\documentclass[conference]{IEEEtran}
\usepackage[numbers,sort]{natbib}
\usepackage{bm}
\usepackage[hidelinks]{hyperref}
\usepackage[names,dvipsnames]{xcolor}
\usepackage{xspace}
\usepackage{amsmath,amsthm,amssymb}
\usepackage{mathtools}
\usepackage{cleveref}
\usepackage{listings}
\usepackage{cprotect}
\usepackage{subfigure}
\usepackage{enumitem}

\usepackage{tikz}
\usetikzlibrary{positioning,shadows,arrows,calc,backgrounds,fit,shapes,snakes,shapes.multipart,decorations.pathreplacing,shapes.misc,patterns}
\tikzset{myiff/.style={double,implies-implies,thick,double equal sign distance}}
\tikzset{myimpl/.style={,double,-implies,double equal sign distance}}
\usepackage{tikzscale}

\usepackage{stmaryrd}
\usepackage[colorinlistoftodos]{todonotes}

\definecolor{codeBlue}{rgb}{0,0,0.85}
\definecolor{orangeGold}{rgb}{0.72, 0.4, 0}
\lstdefinelanguage{sapic}{
  keywords=[1]{
    let, get, in, then, else, suchthat, find,
  },
  keywords=[2]{bitstring,pkey,skey,time,channel,counter_t,key_t,},
  keywords=[3]{foreach, yield, run, defined},
  keywords=[4]{free, type, fun, reduc, equation, end,
    query, equivalence, lemma, theory, begin, builtins, functions, equations, +, lemma, process, table,
    exists-trace,},
  keywords=[5]{All, Ex, forall},
  keywords=[6]{attacker, mess,K,KU,KD},
  keywords=[7]{Accept,Honest},
  sensitive=true,
  morecomment=[n][\itshape]{(*}{*)},
  morecomment=[n][\bfseries]{(**}{*)},
  literate=
    {\&}{{\textsf{\&}}}1
    {:=}{{$\coloneqq$}}2
}

\lstdefinestyle{sapic}{
  language={sapic},
  basicstyle = \footnotesize\ttfamily,
  keywordstyle=[1]{\bfseries\color{codeBlue}},
  keywordstyle=[2]{\color{Plum}},
  keywordstyle=[3]{\bfseries\color{orangeGold}},
  keywordstyle=[4]{\bfseries\color{sqViolet}},
  keywordstyle=[5]{\bfseries\color{sqBrown}},
  keywordstyle=[6]{\bfseries},
  keywordstyle=[7]{\mdseries},
  mathescape = false,
  columns = fullflexible,
  keepspaces = true,
}

\theoremstyle{plain}
\newtheorem{theorem}{Theorem}
\newtheorem{lemma}{Lemma}
\newtheorem{prop}{Proposition}
\newtheorem{axiom}{Axiom}

\theoremstyle{definition}
\newtheorem{defn}{Definition}

\newcommand{\citeposessive}[1]{\citeauthor{#1}'s~[\citenum{#1}]}

\newcommand{\mi}[1]{\ensuremath{\mathit{#1}}}
\newcommand{\mf}[1]{\ensuremath{\mathbf{#1}}}
\newcommand{\ms}[1]{\ensuremath{\mathsf{#1}}}

\newcommand{\srccol}{RoyalBlue}
\newcommand{\trgcol}{RedOrange}
\newcommand{\src}[1]{\mi{\color{\srccol}#1}}
\newcommand{\trg}[1]{\mf{\color{\trgcol}#1}}

\newcommand{\UC}{\ms{UC}\xspace}
\newcommand{\uc}{\textsc{uc}}
\newcommand{\RC}{\mi{RC}\xspace}

\newcommand\PPT{\textrm{\upshape PPT}\xspace}
\newcommand{\NN}{\mathbb{N}}

\newcommand{\env}{\mathcal{Z}}

\newcommand{\Cryptoverif}{\textsc{CryptoVerif}\xspace}
\newcommand{\Wireguard}{\text{Wireguard}\xspace}

\newcommand{\compHyper}{\mi{CH}\xspace}
\newcommand{\hp}{\mi{H}\xspace}

\newcommand{\isdef}{\ensuremath{\mathrel{\overset{\makebox[0pt]{\mbox{\normalfont\tiny\sffamily def}}}{=}}}}
\newcommand{\piff}{\ensuremath{\text{ iff }}}
\newcommand{\comp}[1]{{\color{black}\left\llbracket {#1} \right\rrbracket}}

\DeclareMathOperator{\vdashucperf}{\vdash_{\mkern-4mu\uc}^{=}}
\DeclareMathOperator{\vdashuccomp}{\vdash_{\mkern-4mu\uc}^{\approx}}
\newcommand{\perfuc}[2]{\ensuremath{#1 \vdashucperf #2}}
\newcommand{\compuc}[2]{\ensuremath{#1 \vdashuccomp #2}}

\newcommand{\protocolletter}{\pi} 
\newcommand{\prot}[1]{\trg{\bm{\protocolletter}\ifx\relax#1\relax\else_{#1}\fi}\xspace}

\newcommand{\idealfuncletter}{F}
\newcommand{\idfu}[1]{\src{\idealfuncletter\ifx\relax#1\relax\else_{#1}\fi}\xspace}

\newcommand{\simulatorletter}{S}
\newcommand{\simu}[1]{\src{\simulatorletter\ifx\relax#1\relax\else_{#1}\fi}\xspace}

\newcommand{\attackerletter}{A}
\newcommand{\patt}[1]{\trg{\attackerletter\ifx\relax#1\relax\else_{#1}\fi}\xspace}

\renewcommand{\S}{\src{S}\xspace}
\newcommand{\T}{\trg{T}\xspace}

\newcommand{\actsymbol}{\tau}
\newcommand{\actseq}{\ensuremath{\overline{\actsymbol}}}
\newcommand{\prob}{\ensuremath{\rho}}
\newcommand{\trace}{\ensuremath{\overline{t}}}
\newcommand{\inprefix}{\ensuremath{\overline{\mu}}}

\newcommand{\prefixsymbol}{\ensuremath{m}}
\newcommand{\prefix}{\ensuremath{\overline{\prefixsymbol}}}

\newcommand{\sem}{\mathrel{\rightsquigarrow}}
\newcommand{\sems}{\mathrel{\src{\sem}}}
\newcommand{\semt}{\mathrel{\trg{\boldsymbol{\sem}}}}
\newcommand{\indist}{\approx}
\newcommand{\nindist}{\not\indist}
\newcommand{\compind}{\approxeq}
\newcommand{\statind}{\cong}

\newcommand{\Prop}{\mathcal{T}}
\newcommand{\TT}{\mathbb{T}}

\newcommand{\True}{\mathsf{True}}

\newcommand{\HypEq}[2]{\mathit{Hyp}_{#1,#2}}
\newcommand{\HypEqCls}[2]{\mathit{HypCls}_{#1,#2}}
\newcommand{\EmulN}{\mathit{Emul}}
\newcommand{\Emul}[3]{\EmulN_{{#1},{#2},{#3}}}

\newcommand{\behavname}[2][]{Behav\ifx\relax\detokenize{#1}\relax\else_{#1}\fi(#2)}
\newcommand{\behav}[2][]{\mathit{\behavname[#1]{#2}}}
\newcommand{\behavs}[2][]{\src{\behavname[#1]{#2}}}
\newcommand{\behavt}[2][]{\trg{\behavname[#1]{#2}}}

\newcommand{\programletter}{P}
\newcommand{\prggen}[2]{\programletter\if\relax\detokenize{#1}\relax\else_{#1}\fi\if\relax\detokenize{#2}\relax\else^{#2}\fi\xspace}
\newcommand{\prg}[1]{\mi{\prggen{#1}{}}\xspace}
\newcommand{\prgs}[1]{\src{\prggen{#1}{}}\xspace}
\newcommand{\prgt}[1]{\trg{\prggen{#1}{}}\xspace}

\newcommand{\contextletter}{A}
\newcommand{\ctx}[1]{\contextletter\if\relax\detokenize{#1}\relax\else_{#1}\fi}
\newcommand{\ctxs}[1]{\src{\ctx{#1}}\xspace}
\newcommand{\ctxt}[1]{\trg{\ctx{#1}}\xspace}
\newcommand{\ctxc}{\ctx{}}
\newcommand{\ctxtdummy}{\ctxt{d}}

\newcommand{\linksymbol}[0]{\raisebox{0.33ex}{\ensuremath{\scriptstyle\bowtie}}}
\newcommand{\link}{\mathbin{\linksymbol}}
\newcommand{\linkt}{\mathbin{\trg{\boldsymbol{\linksymbol}}}}
\newcommand{\links}{\mathbin{\src{\linksymbol}}}

\newcommand{\polybase}{poly}
\newcommand{\polyN}{\mi{\polybase}}
\newcommand{\polysN}{\src{\polybase}}
\newcommand{\polytN}{\trg{\polybase}}
\newcommand{\poly}[1]{\polyN(#1)}
\newcommand{\polys}[1]{\src{\polybase(#1)}}
\newcommand{\polyt}[1]{\trg{\polybase(#1)}}

\newcommand{\Predletter}{Q}
\newcommand{\Pred}{\mi{\Predletter}}
\newcommand{\Preds}{\src{\Predletter}}
\newcommand{\Predt}{\trg{\Predletter}}
\newcommand{\Predsa}[1]{\src{\Predletter(#1)}}
\newcommand{\Predta}[1]{\trg{\Predletter(#1)}}

\newcommand{\Exec}{\textsc{Exec}}
\newcommand{\ExecT}{\textsc{ExecT}}

\newcommand{\Execfun}[1]{\Exec\left(#1\right)}

\newcommand{\formatCompilers}[1]{\mi{#1}\xspace}
\newcounter{criteria}
\crefname{criteria}{}{}
\newcommand{\criteria}[2]{%
  \def\thecriteria{\detokenize{#1}}%
    \refstepcounter{criteria}%
    \label{cr:#2}%
    #1%
}

\newcommand{\rhccomp}[0]{\formatCompilers{RHC}}
\newcommand{\rhclabel}[0]{rhc}
\newcommand{\rhcref}[0]{\Cref{cr:\rhclabel}\xspace}
\newcommand{\rhc}[0]{\rhcref}
\newcommand{\rhcdef}[0]{\criteria{\rhccomp}{\rhclabel}}

\newcommand{\comprhccomp}[0]{\formatCompilers{CRHC}}
\newcommand{\comprhclabel}[0]{comprhc}
\newcommand{\comprhcref}[0]{\Cref{cr:\comprhclabel}\xspace}
\newcommand{\comprhc}[0]{\comprhcref}
\newcommand{\comprhcdef}[0]{\criteria{\comprhccomp}{\comprhclabel}}

\newcommand{\predrhccomp}[0]{\formatCompilers{PRHC}}
\newcommand{\predrhclabel}[0]{predrhc}
\newcommand{\predrhcref}[0]{\Cref{cr:\predrhclabel}\xspace}
\newcommand{\predrhc}[0]{\predrhcref}
\newcommand{\predrhcdef}[0]{\criteria{\predrhccomp}{\predrhclabel}}

\newcommand{\predrhpcomp}[0]{\formatCompilers{PRHP}}
\newcommand{\predrhplabel}[0]{predrhp}
\newcommand{\predrhpref}[0]{\Cref{cr:\predrhplabel}\xspace}
\newcommand{\predrhp}[0]{\predrhpref}
\newcommand{\predrhpdef}[0]{\criteria{\predrhpcomp}{\predrhplabel}}

\title{Computationally Bounded Robust Compilation \\ and Universally Composable Security}

\newcommand{\email}[1]{\href{mailto:#1}{\upshape\detokenize{#1}}}
\IEEEoverridecommandlockouts
\author{%
  \IEEEauthorblockN{Robert K\"{u}nnemann}
  \IEEEauthorblockA{CISPA Helmholtz Center for Information Security \\ \email{robert.kuennemann@cispa.de}}
  \and
  \IEEEauthorblockN{Marco Patrignani}
  \IEEEauthorblockA{University of Trento \\ \email{marco.patrignani@unitn.it}}
  \and
  \IEEEauthorblockN{Ethan Cecchetti}
  \IEEEauthorblockA{University of Wisconsin--Madison$^*$\thanks{$^*$Work done in part while author was at the University of Maryland.} \\ \email{cecchetti@wisc.edu}}
}

\begin{document}

\maketitle

\begin{abstract}%
Universal Composability (\UC) is the gold standard for cryptographic security,
but mechanizing proofs of \UC is notoriously difficult.
A recently-discovered connection between \UC and Robust Compilation (\RC)---a novel theory of secure compilation---provides
a means to verify \UC proofs using tools that mechanize equality results.
Unfortunately, the existing methods apply only to perfect \UC security,
and real-world protocols relying on cryptography are only \emph{computationally} secure.

This paper addresses this gap by lifting the connection between \UC and \RC to the computational setting,
extending techniques from the \RC setting to apply to computational \UC security.
Moreover, it further generalizes the \UC--\RC connection beyond computational security to arbitrary equalities,
providing a framework to subsume the existing perfect case, and to instantiate future theories with more complex notions of security.
This connection allows the use of tools for proofs of computational indistinguishability to properly mechanize proofs of computational \UC security.
We demonstrate this power by using \Cryptoverif to  mechanize a proof that parts of the \Wireguard protocol are computationally \UC secure.
Finally, all proofs of the framework itself are verified in Isabelle/HOL.
\end{abstract}

\section{Introduction}
\label{sec:intro}

In cryptography, universal composability (\UC)~\citep{Canetti-UC01} is a framework for the specification and analysis of cryptographic protocols with a key guarantee about compositionality~\citep{simpleUC16,iuc19,ilc19}:
If a protocol is \UC-secure, it behaves like some high-level, secure-by-construction ideal functionality no matter what the protocol interacts with.
If that protocol is then used as a building block inside a larger protocol,
it is safe to replace the smaller protocol with its ideal functionality when reasoning about the security of the larger protocol.
In other words, \UC protocols are secure even when composed with larger protocols.

Proving that a protocol attains \UC security is notoriously complex and error-prone, so it is important to provide ways to mechanize these proofs.
\citet{uc-is-rc22} recently identified a simple and scalable way to define proofs of \UC by relying on a surprising connection between \UC and Robust Compilation~(\RC).

\RC is a hierarchy of criteria for secure compilation introduced by \citet{journey-beyond19,AbateBC+21}.
The criteria describe the security of a compiler by which (hyper)properties~\citep{hyperproperties10} it preserves.

\citet{uc-is-rc22} identify that \UC security is deeply connected to Robust Hyperproperty-Preserving Compilation~(\rhc), the \RC requirement that a compiler preserve arbitrary hyperproperties.
Unfortunately, the connection they identify considers only \emph{perfect} \UC security, where the protocol and ideal functionality exhibit \emph{identical} behaviors.
Real cryptographic protocols are almost never perfectly secure, they rely on cryptographic primitives whose security depends on computational hardness assumptions.
As a result, real ``\UC-secure'' protocols rely on a \emph{computational} definition of \UC security, which allows the protocol to behave differently from the ideal functionality, but only in ways that are indistinguishable to a computationally bounded adversary.
Since the results of \citet{uc-is-rc22} apply only to the perfect case, they provide little help in verifying the far more prevalent proofs of computational \UC security.

For example, consider the following single-bit commitment protocol due to \citet{CanettiF01}.
To commit to bit~$b$, generate a $4n$-bit pseudo-random value~$\mathit{pr}$, and output $\mathit{pr}$ if $b = 0$ and $\mathit{pr} \oplus \sigma$ if $b = 1$ where $\sigma$ is a public $4n$-bit truly random value.
This protocol does not perfectly \UC emulate a functionality that simply indicates a bit has been committed by does nothing else before opening.
An \emph{unbounded} attacker can simply check if the commitment is one of the (exponentially many) possible pseudo-random values and correctly guess the value of $b$ with overwhelming probability.
The protocol is, however, \emph{computationally} \UC secure given a pseudo-random generator with the right structure~\citep{CanettiF01},
a proof that relies on a \emph{polynomial} adversary's inability to distinguish~$\mathit{pr}$ from a truly random value.

This work addresses the limitation of \citet{uc-is-rc22} by lifting their results to the computational case,
allowing us to consider the security of protocols like \citeauthor{CanettiF01}'s commitment.
To accomplish this goal, we replace equality of behaviors with \emph{computational indistinguishability} in both the \UC and \RC theories.
Making this switch in both the \UC and \RC contexts produces a notion of robust compilation that corresponds precisely to the already-established definition of computational \UC security.

On the \RC side, the change requires two fundamental modifications.
First, \citet{uc-is-rc22} describe programs as producing probability distributions over possible traces and consider a protocol (or compiler) secure if the ideal functionality (source program) and protocol (compiled program) produce identical distributions.
Computational indistinguishability, however, does not relate individual distributions.
It relates \emph{families} of distributions~$\{X_n\}$ and~$\{Y_n\}$ indexed by a security parameter~$n$,
and requires an adversary's ability to distinguish between~$X_n$ and~$Y_n$ to shrink quickly as~$n$ grows.
We thus expand the definition of program behavior to include explicit security parameters.
Second, \RC definitions in general define robustness for all programs against all adversaries, while computational \UC-security concerns only \emph{polynomial-time} programs and attackers.
We therefore extend the \RC framework to consider specific classes of protocols and attackers.

With these two modifications to the \RC theory of \citet{journey-beyond19},
we can define a new class of hyperproperties,~\compHyper, where each hyperproperty is the set of behaviors that are computationally indistinguishable from some ideal functionality.
We also define a notion of Computationally-Robust Hyperproperty-Preserving Compilation~(\comprhc), a new notion of \RC that preserves \compHyper against polynomial-time attackers.
Finally, we prove that computational \UC security is equivalent to \comprhc.

Technically, proving \comprhc amounts to proving that some source program (an ideal functionality), linked with an existentially bound attacker (a simulator) is computationally indistinguishable from a target program (a protocol).
Fortunately, existing tools such as \Cryptoverif~\citep{cryptoverif} provide ways to mechanize such proofs.
Thus, we showcase the ability to provide scalable proofs of computational \UC via proofs of \comprhc by using \Cryptoverif to mechanize a proof of computational indistinguishability for the \Wireguard protocol~\citep{lippMechanisedCryptographicProof2019}.

Notably, the results of lifting the connection between \UC and \RC to the computational case contain nothing specific about computational indistinguishability and polynomial time.
As a result, we are able to substantially generalize the theory
by using any equivalence~$\equiv$ defining indistinguishable behaviors,
and an arbitrary predicate~$\Pred$ on programs and contexts to define the class of programs and contexts.
Doing so produces a notion of \UC security up-to~$\equiv$, an \RC notion of $\Pred$-robust preservation of $\equiv$-hyperproperties, and a proof that the two are equivalent.
This result immediately subsumes the original connection of \citet{uc-is-rc22}, using $=$ as $\equiv$ and the trivial predicates allowing all programs and contexts,
our lifted result, using computational indistinguishability as the equivalence and polynomial time as the predicate,
and suggests more definitions and connections to explore.

To summarize, the main contributions of this paper are:
\begin{itemize}
  \item \Cref{sec:comp-rc} explicitly models security parameters and computational indistinguishability in the \RC framework, and uses these structures to extend the result of \citet{uc-is-rc22} to computational security.
  \item \Cref{sec:general-equivs} generalizes the computational result to arbitrary indistinguishability relations, making perfect security, computational security, and many other interesting equivalences special cases of a general theorem.
  \item \Cref{sec:wireguard} uses these results to mechanize the proof of \UC for the WireGuard protocol using the \Cryptoverif tool, which provides computational security guarantees.
\end{itemize}
The rest of this paper provides background notions related to the main results (\Cref{sec:background}), 
interesting details of the proofs of our main theorems (\Cref{sec:proof-approach}),
the presentation of related work (\Cref{sec:rw}), and conclusions (\Cref{sec:conc}).

All theorems in Sections~\ref{sec:comp-rc},~\ref{sec:general-equivs}, and~\ref{sec:proof-approach} are verified in the Isabelle/HOL theorem prover~\citep{isabelle},
while theorems in \Cref{sec:wireguard} are verified in \Cryptoverif~\citep{cryptoverif}.
The proof development is available at: \url{https://uc-is-sc.github.io/}.

\section{Background}
\label{sec:background}

This section presents relevant background notions on Universal Composability (\Cref{sec:uc-background}), Robust Compilation (\Cref{sec:rc-background}) and their connection (\Cref{sec:existing-ucrc-conn}).

For aide reading~\cite{patrignani2020use}, throughout the paper we will use $\src{blue~italic}$ for functionalities and source programs,
$\trg{bold~red}$ for protocols and target programs,
and black for terms not specific to either context.

\subsection{Universally Composable Security}
\label{sec:uc-background}

Universally Composable~(\UC) security~\citep{Canetti-UC01}
defines a security notion that combines functional correctness and
privacy and is both transitive as well as closed under protocol
composition. This is achieved by refinement: a ``secure'' protocol is
one that is
``at least as secure'' as a protocol where all parties forward their
communication to a single entity that 
(a) computes the correct output (which they forward to the environment)
and
(b) leaks only the minimal amount of information via the network
(which is conservatively modelled by sending this information to the attacker).
There is a distinction between the
environment, which models the trusted input and output from
either higher-level protocols or the user of the system,
and
the attacker, which models the hostile network.
Together, 
the environment~$\env$,
protocol~$\prot{}$
and
attacker~$\patt{}$
constitute a system
that can be executed.
The execution ends when the environment decides whether it is interacting with the real protocol or an ideal simulation
and outputs a final bit to indicate its guess.
Let $\Exec(\env, \patt{}, \prot{})$ be a random variable describing the outcome of this probabilistic process.

The strongest notion of ``at least as secure as'' says that any attack on a protocol can be simulated using the functionality.
That is, only knowing the ``acceptable'' leakage built into the functionality is enough to convincingly reproduce any real attacker's behavior.
For an encryption functionality, for instance, this leakage is the message length.

If this simulation can always produce \emph{identical} behavior to the real attacker,
the protocol is said to \emph{perfectly} emulate the ideal functionality.
\begin{defn}[Perfect \UC Emulation]\label{def:perf-uc}
  A protocol~$\prot{}$ \emph{perfectly \UC-emulates} a functionality~$\idfu{}$, denoted \perfuc{\prot{}}{\idfu{}}
  if, for all (unbounded) adversaries~$\patt{}$, there is a simulator~$\simu{}$, such that for all environments~$\env$,
  $$\Exec(\env, \patt{}, \prot{}) = \Exec(\env, \simu{}, \idfu{})$$
\end{defn}

For realistic cryptographic protocols, however, perfect emulation is impossible.
Their security almost invariably relies on computational hardness assumptions, so an unbounded attacker or environment
can easily glean information beyond the ideal functionality's ``acceptable'' leakage.
As a result, the random variables will not be identically distributed.

To still recover a meaningful notion of security, we include a security parameter~$n$
and compare the behaviors asymptotically in~$n$.
That is, instead of considering $\Exec(\env, \patt{}, \prot{})$ to be a single random variable,
we consider it to be a family of random variables, $\{\Exec_n(\env, \patt{}, \prot{})\}$, one for each value of~$n$.
Two such families are \emph{indistinguishable} if the difference between them shrinks very rapidly (usually exponentially) as~$n$ grows.
This notion is defined formally as follows.

\begin{defn}[Indistinguishability \citep{Canetti-UC01}]
  \label{def:indist}
  Two ensembles of binary probability distributions $X = \{X_n\}$ and $Y = \{Y_n\}$
  are \emph{indistinguishable}, denoted $X \approx Y$, if,
  for all $c \in \NN$, there is some $N \in \NN$ such that
  $$\forall n > N.\, | \Pr[X_n = 1] - \Pr[Y_n = 1] | < n^{-c}$$
\end{defn}

Indistinguishability is not enough by itself.
The computational hardness assumptions of the cryptographic primitives force us to limit the computational power of the parties within the system.
In particular, the \UC framework demands that all protocols, attackers, and environments execute in polynomial time.
The result is the following formal definition of computational \UC security.

\begin{defn}[Computational \UC Emulation~\citep{Canetti-UC01}]\label{def:comp-uc}
  A poly-time protocol $\prot{}$ \emph{computationally \UC-emulates} a functionality~$\idfu{}$, denoted \compuc{\prot{}}{\idfu{}}
  if, for all \PPT adversaries~$\patt{}$, there is a \PPT simulator~$\simu{}$, such that for all \PPT environments~$\env$,
  $$\Exec(\env, \patt{}, \prot{}) \approx \Exec(\env, \simu{}, \idfu{})$$
\end{defn}

\subsection{Robust Compilation}
\label{sec:rc-background}

The Robust Compilation (\RC) framework~\citep{journey-beyond19,AbateBC+21} formalizes security of compilers as their ability to preserve arbitrary classes of (hyper)properties~\citep{hyperproperties10} \emph{robustly}, i.e., in the presence of active adversaries linked with compiled code.

Compilers (denoted by $\comp{\cdot}$) are functions that translate components, or \emph{partial programs} ($\prg{}$), from a source language~(\S) to a target language~(\T).
In both languages, a partial program~$\prg{}$ can link with a \emph{program context}~$\ctxc{}$ to form a \emph{whole program}, denoted~$\ctxc{}\link\prg{}$.
Whole programs come equipped with an operational semantics called a \emph{robust trace semantics}~($\sem$),
which captures all security-relevant behavior of $\ctxc{} \link \prg{}$ in a trace of events~($\trace$)
where each event indicates whether it comes from the context or the program.

In their original work, \citet{journey-beyond19} define what criteria compilers must meet to prove that they preserve:
\begin{itemize}
  \item trace properties, including all trace properties, safety properties, and dense properties, a variation of liveness,
  \item hyperproperties, including all hyperproperties, subset-closed hyperproperties, and 2-hypersafety, and
  \item relational hyperproperties.
\end{itemize}
A compiler that robustly preserves all hyperproperties satisfies \rhcdef, defined formally as follows.
\begin{defn}[Robust Hyperproperty-Preserving Compiler]\label{def:rhc}
  \begin{align*}
    \vdash \comp{\cdot} : \rhc \isdef
    \forall \prgs{}, \ctxt{}\ldotp \exists \ctxs{}\ldotp \forall \trace\ldotp \ctxt{}\linkt{}\comp{\prgs{}} \semt \trace \piff \ctxs{}\links{}\prgs{} \sems \trace
  \end{align*}
\end{defn}
For a compiler to preserve arbitrary hyperproperties, the traces exhibited by any compiled component while interacting with an arbitrary target context must be the same to the traces exhibited by a the source component linked with an existentially quantified source context.

\subsection{Existing \UC--\RC Connection}
\label{sec:existing-ucrc-conn}

As \citet{uc-is-rc22} prove, \Cref{def:perf-uc} and \Cref{def:rhc} are equivalent.
By inspecting the quantifiers, and their order (which match), it is possible to see that \UC functionalities correspond to \RC source components, \UC protocols correspond to \RC compiled components, and \UC environments correspond to \RC traces.
The elements that provide a central role in the security argument are also related.
The universally quantified \UC protocol attackers are still universally quantified \RC target contexts and the existentially quantified \UC simulators are still existentially quantified \RC source contexts.

With this equivalence, \citet{uc-is-rc22} demonstrate that by providing an \rhc proof for a compiler that translates a program encoding a functionality into its protocol, one can obtain a \UC proof.
More importantly, since \Cref{def:rhc} amounts to proving trace equivalence between programs, \citet{uc-is-rc22} use the Deepsec tool~\citep{chevalDEEPSECDecidingEquivalence2018} in order to provide the \rhc proof and thus the first mechanized proof of \UC.

However, the equivalence formalized by \citet{uc-is-rc22} only amounts to a perfect notion of \UC.
In the following section, we set out to lift this limitation.

\section{Computational Robust Hyperproperty Preservation}
\label{sec:comp-rc}

To see the connection between perfect \UC and \rhc,
we look at the relationship between the definition of \UC security and the behavior of a program~$W$, denoted $\behav{W}$.
\citet{uc-is-rc22} define $\behav{W} = \{ \trace \mid W \sem \trace\}$ as the set of traces~$W$ can produce,
and a trace $\trace$ as a pair $(\actseq, \prob)$, where $\actseq$ is a (potentially infinite) sequence of actions
and $\prob$ is the probability that~$W$ will produce~$\actseq$.
They also include all actions of the environment in~$\actseq$ (including outputting the final bit),
so we can view $\behav{W}$ as containing different probability distributions for different environments.
If we let $\behav{W}|_\env$ denote the probability distribution over final bits
for just the subset of behaviors consistent with environment~$\env$,
the structure of \rhc begins to look very much like the structure of \UC security.

Indeed, we can rephrase \Cref{def:rhc} to require equality of behaviors
and then universally quantify over environments.
The result is the following definition of \rhc.
$$\forall \prgs{}, \ctxt{}\ldotp \exists \ctxs{}\ldotp \forall \env\ldotp \behavt{\ctxt{}\linkt\comp{\prgs{}}}|_\env = \behavs{\ctxs{}\links\prgs{}}|_\env$$
And recall from \Cref{def:perf-uc} that $\prot{}$ perfectly \UC-emulates~$\idfu{}$~if
$$\forall \patt{}\ldotp \exists \simu{}\ldotp \forall \env\ldotp \Exec(\env, \patt{}, \prot{}) = \Exec(\env, \simu{}, \idfu{})$$

These definitions have a nearly-identical structure, which we use as a guide for moving to computational security.
There are two main differences between perfect \UC and computational \UC that we need to represent in the language of robust compilation.
First, the equality on distributions is replaced with computational indistinguishability.
Second, the relationship between real and ideal protocols (and thus target and source programs, respectively)
is further loosened to consider only attackers, simulators, and programs that are \emph{polynomial time}.

This section now presents these changes to the existing \RC theory (\Cref{sec:comp-rc:ind}) followed by the novel notion of computational robust compilation (\comprhc, \Cref{sec:comp-rc:poly}).
Then, it presents what class of hyperproperties \comprhc preserves (\Cref{sec:comp-rc:hyperprops}) before presenting the main result of this paper, connecting omputational \UC and \comprhc (\Cref{sec:comp-rc:uc}).

\subsection{Computational Indistinguishability}
\label{sec:comp-rc:ind}

Defining indistinguishability of program behaviors requires modifying the structures introduced by \citet{uc-is-rc22}.
Recall from \Cref{sec:uc-background} that indistinguishability, denoted \mbox{$X \approx Y$},
is defined over \emph{families} of distributions, $X = \{X_n\}$ and $Y = \{Y_n\}$,
and requires the ability of an adversary to separate $X_n$ from $Y_n$ to shrink rapidly as~$n$ grows.
To modify the above definition of \rhc to match the structure of computational \UC security,
we would like to say \mbox{$\behav{W_{\rm 1}}|_\env \approx \behav{W_{\rm 2}}|_\env$}.

Unfortunately, with $\behav{W}$ structured as a set of pairs~$(\actseq, \prob)$, as described above,
we can interpret $\behav{W}|_\env$ as a single probability distribution, but not the requisite family of them.
Luckily, because the semantics of the program are left abstract, we can straightforwardly introduce a security parameter~$n$.
We thus define a program execution by a triple~$(\actseq, \prob, n)$ instead of a pair.
Here~$\actseq$ and~$\prob$ are the same as before, and $n$ is a security parameter.
With this change, we can view $\behav{W}$ as a \emph{family of behaviors}, indexed by~$n$,
which we can similarly restrict by an environment to get a family of distributions.
That is,
\begin{align*}
  \behav[n]{W} & \isdef \{ (\actseq, \prob) \mid W \sem (\actseq, \prob, n) \} \\
  \behav{W} & \isdef \{ \behav[n]{W} \} \\
  \behav{W}|_\env & \isdef \{ \behav[n]{W}|_\env \}
\end{align*}
\Cref{def:indist} now applies with these families of distributions.

Quantifying over all possible environments, however, results in a definition of statistical indistinguishability, not computational indistinguishability.
Achieving the computational goal requires restricting to only polynomial-time environments.
Because we assume the program interfaces of the different languages match, the space of environments is the same,
we use the same $\PPT$ set to represent all probabilistic poly-time environments.

These two modifications are sufficient to define computationally indistinguishable program behaviors.
\begin{defn}[Computational Indistinguishability of Programs]
  \label{def:prgm-comp-ind}
  Whole programs $W_1$ and $W_2$ have \emph{computationally indistinguishable behavior} if
  $$\forall \env \in \PPT\ldotp \behav{W_{\rm 1}}|_\env \approx \behav{W_{\rm 2}}|_\env$$
  We denote this equivalence by $\compind$ for both whole programs and sets of traces.
  That is, the above equivalence defines both $W_1 \compind W_2$ and $\behav{W_{\rm 1}} \compind \behav{W_{\rm 2}}$.
\end{defn}

\subsection{Computational Robust Compilation}\label{sec:comp-rcc}
\label{sec:comp-rc:poly}

Recall that computational \UC security allows not only this looser computational notion of indistinguishability,
but also restricts all protocols, attackers, and simulators to be polynomial time.
The language of robust compilation, however, quantifies over \emph{all} programs and contexts, not just poly-time ones.

Without bounding the computational power of programs and contexts, we cannot hope to represent computational \UC security.
To see why, consider the single-bit commitment protocol due to \citet{CanettiF01} discussed in \Cref{sec:intro}.
In the \RC context, if we consider a compiler that takes \citeauthor{CanettiF01}'s commitment ideal functionality
and produces their commitment protocol, we would hope to call that compiler robust.
But because the language of \RC always considers \emph{all} contexts (attackers) and a computationally unbounded attacker can break the protocol, we cannot.
To address this concern, we expand the framework with a notion of \emph{computationally-robust} preservation of hyperproperties
that considers only polynomial-time programs and contexts.

As in \UC, execution complexity depends on the context (adversary), the program (protocol), and how they interact.
We therefore define a poly-time predicate over whole programs.
As with the program semantics, we leave this predicate abstract to avoid the need to restrict
to a specific computational model with specific execution times.
Because the source and target languages may have different semantics and computational models,
we include both a source-language predicate $\polysN$ and a target-language predicate $\polytN$.

These predicates are sufficient to define a more permissive notion of robust compilation:
the \emph{computationally-robust hyperproperty-preserving compiler}~(\comprhcdef).
\begin{defn}[Computationally-Robust Hyperproperty-Preservating Compiler]\label{defn:comprhc}
  \begin{align*}
    \vdash \comp{\cdot} : \comprhc \isdef{} & \forall \prgs{}\ldotp \forall \ctxt{}\ldotp \polyt{\ctxt{}\linkt\comp{\prgs{}}} \Longrightarrow \\
    & \quad \exists \ctxs{}\ldotp \polys{\ctxs{}\links\prgs{}} \land \left(\ctxt{}\linkt\comp{\prgs{}} \compind \ctxs{}\links\prgs{}\right)
  \end{align*}
\end{defn}

This computational notion considers only executions of poly-time programs in both the source and target language.
It demands that, for any target context~$\ctxt{}$ where $\ctxt{}\linkt\comp{\prgs{}}$ is poly-time,
there must be a source context~$\ctxs{}$ such that $\ctxs{}\links\prgs{}$ is also poly-time
and the behaviors of the whole programs are computationally indistinguishable.

\paragraph*{\comprhc and Optimizations}
\comprhc demands that, if there is \emph{any} target context~$\ctxt{}$ such that $\ctxt{}\linkt\comp{\prgs{}}$
is poly-time, then there must be some source context~$\ctxs{}$ such that $\ctxs{}\links\prgs{}$ is also poly-time.
This requirement has an odd ramification:
a \comprhc compiler may not optimize a super-polynomial program into a poly-time one, say, by introducing memoization.
This may appear to be a limitation, but from a security standpoint it is not.
Computational \UC security considers only the behavior of poly-time programs and poly-time ideal functionalities.
It says nothing about the security of a system with super-polynomial protocols, adversaries, environments, or ideal functionalities.
Correspondingly, the security defined by \comprhc does not aim to be meaningful for super-polynomial programs.
Instead, it demands that the compiler produce super-polynomial outputs on any super-polynomial inputs,
allowing us to restrict all relevant analysis to the polynomial case.

\subsection{Identifying Preserved Hyperproperties}
\label{sec:comp-rc:hyperprops}

In the original work defining \rhc, \citet{journey-beyond19} identify that \rhc is equivalent to a compiler robustly preserving all hyperproperties.
Recall that a hyperproperty is a set of sets of traces.
This means that, for $\comp{\cdot}$ to be \rhc,
then for any hyperproperty~$\hp$ and partial program~$\prgs{}$,
if the behavior of $\prgs{}$ is in $\hp$ for every possible source context,
then the behavior of $\comp{\prgs{}}$ is in $\hp$ for every possible target context.

More formally, let $\Prop = \{ \trace \}$ be the set of all traces and $\TT = \mathcal{P}(\Prop)$ be the set of sets of traces.
Then
\begin{align*}
  \vdash \comp{\cdot} :{} & \rhc \iff \\
  & \forall \hp \subseteq \TT\ldotp \forall \prgs{}\ldotp
  \begin{array}[t]{@{}l@{}}
    (\forall \ctxs{}\ldotp \behavs{\ctxs{}\links\prgs{}} \in \hp) \Longrightarrow \\
    (\forall \ctxt{}\ldotp \behavt{\ctxt{}\linkt\comp{\prgs{}}} \in \hp)
  \end{array}
\end{align*}

This connection with \rhc raises a question in our context:
what hyperproperties does computational robustness preserve?
Or, more generally speaking: which class of (hyper)properties can \UC security actually express?
One might guess that \comprhc is equivalent to computationally-preserving all hyperproperties.
That is, preserving all hyperproperties when considering only poly-time programs.
However, restricting to poly-time environments, which attempt to distinguish program behaviors, makes this guess incorrect.
Instead, \comprhc corresponds to preserving a narrower class of hyperproperties,~$\compHyper$,
which represent computationally-indistinguishable programs.

To define $\compHyper$, we first define $\compHyper(F)$, a single hyperproperty representing
the family of behaviors that are computationally indistinguishable from what~$\idealfuncletter$
can produce when given  an appropriate context.
Formally,
\begin{align*}
  \compHyper(\idealfuncletter) \isdef \{T \in \TT & \mid \exists \simulatorletter\ldotp \poly{\simulatorletter \link \idealfuncletter} \\
  & \land \left(\forall \env \in \PPT\ldotp T|_\env \approx \behav{\simulatorletter \link \idealfuncletter}|_\env\right) \}
\end{align*}
Intuitively, $\behav{W} \in \compHyper(\idealfuncletter)$ means that there is some simulator~$\simulatorletter$
such that the behavior of $\simulatorletter \link \idealfuncletter$ cannot be distinguished from the behavior of~$W$.
If we instead consider a partial program~$\prg{}$ and quantify over poly-time contexts,
$$\forall \contextletter\ldotp \poly{\contextletter \link \prg{}} \Longrightarrow \behav{\contextletter \link \prg{}} \in \compHyper(\idealfuncletter)$$
means that $\prg{}$ computationally emulates $\idealfuncletter$.

The class $\compHyper$ that a \comprhc compiler preserves (for poly-time programs)
is precisely the set of all of these hyperproperties.
That is, we define
$\compHyper \isdef \{\hp \mid \exists \idealfuncletter\ldotp \hp = \compHyper(\idealfuncletter)\}$
and obtain the following result.
\begin{theorem}[\comprhc is Computational Preservation of $\compHyper$]
  \label{thm:comprhc-ch}
  \begin{align*}
    \vdash\comp{\cdot} :{} & \comprhc \iff \forall \hp \in \compHyper\ldotp \\
    \forall \prgs{}\ldotp &
    \left(\forall \ctxs{}\ldotp \polys{\ctxs{}\links\prgs{}} \Longrightarrow \behavs{\ctxs{}\links\prgs{}} \in \hp\right) \Longrightarrow \\
    & \left(\forall \ctxt{}\ldotp \polyt{\ctxt{}\linkt\comp{\prgs{}}} \Longrightarrow \behavt{\ctxt{}\linkt\comp{\prgs{}}} \in \hp\right)
  \end{align*}
\end{theorem}

This theorem, proven in Isabelle/HOL ($\mathtt{compRHCisRHPX}$),
is a special case of \Cref{thm:predrhc-is-predrhpx} (see \Cref{sec:general-equivs}).

\Cref{thm:comprhc-ch} shows that \comprhc is equivalent to preserving secure emulation.
That is, a compiler $\comp{\cdot}$ satisfies \comprhc if and only if,
whenever a source program $\prgs{}$ computationally emulates some functionality~$\idealfuncletter$ in the source language,
then the compiled $\comp{\prgs{}}$ also computationally emulates~$\idealfuncletter$ in the target language.

\subsection{Connecting Computational \UC and \comprhc}
\label{sec:comp-rc:uc}

This equivalence between \comprhc and secure emulation of functionalities suggests
a similarly deep connection to computational \UC security.
Indeed, the modifications to get from \rhc to \comprhc were designed explicitly to mirror the differences between perfect \UC and computational \UC.
To prove this correspondence formally, we rely on four axioms laid out by \citet{uc-is-rc22}
to move between the Interactive Turing Machine~(ITM) semantics of the \UC framework and the abstract semantics of the \RC framework.
We modify Axioms~\ref{ax:uc-rc}, \ref{ax:canon-env}, and~\ref{ax:final-bit} only to add the security parameter~$n$.

The first uses the function $z(\prefix)$ to define a \emph{canonical environment}
for a trace prefix $\prefix$ that produces exactly this prefix and then halts the execution with final bit 1.
It says that a trace prefix is possible if and only if a corresponding execution is possible for the ITM.
\begin{axiom}[\UC and \RC
    Semantics~\citep{uc-is-rc22}]\label{ax:uc-rc}
  If $\prefix = (\inprefix, \prob, n)$ and $\inprefix$ is a finite sequence of actions, then
  $$\ctx{}\link{\prg{}}\sem \prefix
  \iff \Pr[\ExecT_n(z(\prefix),\ctx{},\prg{}) = \inprefix] = \prob > 0.$$
\end{axiom}

The second axiom requires that the canonical environment for a trace correctly represent the behavior
of all environments that produce the same trace.
\begin{axiom}[Canonical Environment Correctness~\citep{uc-is-rc22}]\label{ax:canon-env}
  If $\env$ is non-probabilistic, $\prefix = (\inprefix, \prob, n)$,
  and for some $\trg{\contextletter'}$ and $\trg{\prot{}'}$, $\Pr[\ExecT_n(\env, \trg{\contextletter'}, \trg{\prot{}'}) = \inprefix] > 0$,
  then for all $\patt{}$ and $\prot{}$,
  $$\Pr[\ExecT_n(\env{},\patt{},\prot{}) = \inprefix ] = \Pr[\ExecT_n(z(\prefix),\patt{},\prot{}) = \inprefix ]$$
\end{axiom}

This axiom assumed that the environment~$\env$ is non-probabilistic, meaning the probability of~$\prefix$
depends only on the randomness of $\patt{}$ and $\prot{}$.
This assumption simplifies reasoning and is not a limitation, because we also assume that
anything a probabilistic environment can do, a non-probabilistic one can do as well.
\begin{axiom}[Non-Probabilistic Environment Completeness]\label{ax:nonprob-complete}
  For any $\env \in \PPT$, if $\Execfun{\env,\patt{},\prot{}} \nindist \Execfun{\env,\simu{},\idfu{}}$,
  then there exists some non-probabilistic poly-time $\env'$ such that
  $\Execfun{\env',\patt{},\prot{}} \nindist \Execfun{\env',\simu{},\idfu{}}$.
\end{axiom}
\citeposessive{uc-is-rc22} version of Axiom~\ref{ax:nonprob-complete} uses (in)equality
instead of (in)distinguishability and does not bound~$\env$ and~$\env'$, but is otherwise the same.
Intuitively, this axiom is valid because, for any distinguishing~$\env$,
one can select from the random choices made by $\env$ (but not the attacker or protocol)
the ones that maximize $\env$'s ability to distinguish the real and ideal worlds.
Fixing those choices produces a deterministic environment that distinguishes at least as well as~$\env$.

Finally, we assume that trace prefixes specify whether or not the environment has decided on a final bit $b$, and if so, what that value is.
The \UC framework assumes the environment terminates execution with that final bit,
but we follow \citet{uc-is-rc22} and abstract this process over an arbitrary encoding with an extraction function $\beta : \inprefix \to \{0, 1, \bot\}$
with the following property.
\begin{axiom}[Finite Traces Contain the Final Bit~\citep{uc-is-rc22}]\label{ax:final-bit}
  \begin{align*}
    \Pr[&\Exec_n(\env,\patt{},\prot{})= b ] \\
    & = \sum_{\beta(\inprefix)=b} \Pr[\ExecT_n(\env,\patt{},\prot{}) = \inprefix]
  \end{align*}
\end{axiom}

Using these axioms, we are able to prove the desired correspondence between \UC security and \comprhc.
\begin{theorem}[Computational \UC and \comprhcref
    Coincide]\label{thm:compRHCeqUC}
  $$\vdash \comp{\cdot} : \comprhc \iff \forall \prgs{}\ldotp \compuc{\comp{\prgs{}}}{\prgs{}}$$
\end{theorem}
\noindent
This theorem is verified in Isabelle/HOL ($\mathtt{compRHCeqUC}$).

As we will see in \Cref{sec:wireguard}, \Cref{thm:compRHCeqUC} creates a powerful new means of mechanizing a proof of \UC. 
It is now sufficient to show that the compiler that translates \idfu{} into \prot{} satisfies \comprhc or, equivalently, preserves \compHyper for polynomial-time programs.
That is, it suffices to prove computational indistinguishability between~\idfu{} and~\prot{}, given the correct simulator~\simu{}.
These proofs can be carried out using existing tools. We will use \Cryptoverif~\cite{cryptoverif} and discuss alternatives in \Cref{sec:rw}.

\section{Generalizing Equivalences and Predicates}
\label{sec:general-equivs}

In \Cref{sec:comp-rc:ind,sec:comp-rc:poly} we modified \rhc to obtain the computational analogue \comprhc.
Very little about that process, however, was specific to polynomial time or computational indistinguishability.
Indeed, we can generalize nearly all of those modifications and achieve a far more general result
that subsumes our results from \Cref{sec:comp-rc}, as well as the perfect security results of \citet{uc-is-rc22},
and points to other notions of security as well.

In \Cref{sec:comp-rc:ind}, we loosened the requirement of \emph{equality} of behaviors to \emph{computational indistinguishability}.
This change required modifying the definition of traces to interpret a set of traces as a family of probability distributions.
To capture both the original equality view and this looser view, we can generalize to an arbitrary equivalence relation $\equiv$ over sets of traces,
which we will apply to program behaviors.

In \Cref{sec:comp-rc:poly}, we expanded the robust compilation framework with polynomial-time predicates.
We required that behaviors only be preserved for poly-time programs,
though we restricted the existentially quantified source contexts to be poly-time as well.
There is nothing special about poly-time in that process.
Indeed, we could instead have used abstract predicates $\Preds$ and $\Predt$ over source and target programs.

Making this change leads to a more general notion of \emph{predicate-robust hyperproperty preservation} (\predrhcdef)
parameterized on an equivalence and two predicates.
\begin{defn}[Predicate-Robust Hyperproperty-Preservating Compiler]
  \label{def:pred-rhc}
  \begin{align*}
    \vdash \comp{\cdot} & : \predrhc(\equiv, \Predt, \Preds) \isdef
      \forall \prgs{}\ldotp \forall \ctxt{}\ldotp \Predta{\ctxt{}\linkt\comp{\prgs{}}} \Longrightarrow
    \\
    & \exists \ctxs{}\ldotp \Predsa{\ctxs{}\links\prgs{}}
      \land \left(\behavt{\ctxt{}\linkt\comp{\prgs{}}} \equiv \behavs{\ctxs{}\links\prgs{}}\right)
  \end{align*}
\end{defn}
That is, source contexts linked with source programs can produce
all the behavior of compiled programs linked with target contexts (up to~$\equiv$)
when only considering programs satisfying $\Preds$ and $\Predt$, the respective predicates for the source and target language.

This notion generalizes both \rhc and \comprhc.
If we instantiate $\equiv$ with set equality ($=$) and both $\Preds$ and $\Predt$ with the trivial predicate $\True$ that holds for all programs,
$\predrhc(=, \True, \True)$ requires source contexts to produce exactly the behavior of target contexts
when considering all programs---precisely \rhc.
\begin{prop}[\predrhc expresses \rhc]
$$\def\arraystretch{1.25}
\begin{array}{c}
  \vdash \comp{\cdot} : \predrhc(=, \True, \True) \\
  \Updownarrow \\
  \forall \prgs{}\ldotp \forall \ctxt{}\ldotp \exists \ctxs{}\ldotp \behavt{\ctxt{}\linkt\comp{\prgs{}}} = \behavs{\ctxs{}\links\prgs{}} \\
\end{array}$$
which is the definition of $\vdash \comp{\cdot} : \rhc$.
\end{prop}

Similarly, we can instantiate $\equiv$ with computational indistinguishability ($\compind$),
interpreting traces as families of distributions and restricting to poly-time environments,
$\Preds$ with $\polysN$, and $\Predt$ with $\polytN$.
Here $\predrhc(\compind, \polysN, \polytN)$ says that source contexts must produce behavior
indistinguishable from target contexts when considering only polynomial programs---precisely \comprhc.
\begin{prop}[\predrhc expresses \comprhc]\label{prop:comprhc-is-predrhc}
  $$\vdash \comp{\cdot} : \predrhc(\compind, \polytN, \polysN) \iff {\vdash \comp{\cdot} : \comprhc}$$
\end{prop}

We also generalize the result from \Cref{sec:comp-rc:hyperprops}
identifying which hyperproperties \comprhc compilers preserve.
To formalize this idea,
we extend robust preservation of a specific class~$X$ of hyperproperties, due to \citet{journey-beyond19},
with predicates in the same way that \predrhc extends \rhc,
producing the following definition of \emph{predicate-robust hyperproperty preservation} (\predrhpdef).

\begin{defn}[Predicate-Robust Hyperproperty Preservation]
  \label{def:pred-rhpx}
  \begin{align*}
    \vdash \comp{\cdot} & : \predrhp(X, \Predt, \Preds) \isdef 
    \forall \hp \in X\ldotp \forall \prgs{}\ldotp \\
    & \quad \left(\forall \ctxs{}\ldotp \Predsa{\ctxs{}\links\prgs{}} \Longrightarrow \behavs{\ctxs{}\links\prgs{}} \in \hp\right) \Longrightarrow \\
    & \quad \left(\forall \ctxt{}\ldotp \Predta{\ctxt{}\linkt\comp{\prgs{}}} \Longrightarrow \behavt{\ctxt{}\linkt\comp{\prgs{}}} \in \hp\right)
  \end{align*}
\end{defn}

That is, a $\predrhp(X, \Predt, \Preds)$ compiler is one where, for any hyperproperty $\hp \in X$,
if a source program~$\prgs{}$ produces only behaviors in~$\hp$ when restricting to source contexts that satisfy~$\Preds$,
then the compiled program $\comp{\prgs{}}$ must also produce only behaviors in~$\hp$
when considering only target contexts satisfying~$\Predt$.

As with \predrhc, this generalizes previous notions.
It can express robust preservation of all hyperproperties through $\predrhp(\mathcal{P}(\TT), \True, \True)$,
setting~$X$ to all hyperproperties and $\Preds$ and $\Predt$ to $\True$ to consider all contexts.
It can also express computationally-robust preservation of~$\compHyper$ (\Cref{sec:comp-rc:hyperprops}),
by setting $X = \compHyper$ and $\Preds$ and $\Predt$ to $\polytN$ and $\polysN$, respectively.

Finally, we can relate \predrhc and \predrhp by
defining the class of hyperproperties that a \predrhc compiler preserves
in terms of the equivalence~$\equiv$ and the predicates~$\Preds$ and~$\Predt$.
As in \Cref{sec:comp-rc:hyperprops}, we define the class of hyperproperties by the set of functionalities~$\idealfuncletter$ that a program securely emulates.
Where $\compHyper(\idealfuncletter)$ is the set of behaviors computationally indistinguishable from~$\idealfuncletter$,
the general hyperproperty $\HypEq{\equiv}{\Pred}(\idealfuncletter)$ is the set of behaviors equivalent to~$\idealfuncletter$ up to~$\equiv$
when considering only whole programs that satisfy predicate~$\Pred$.
Formally,
\begin{align*}
  \HypEq{\equiv}{\Pred}(\idealfuncletter) & \isdef 
  \\
  \{T \in \TT & \mid \exists \simulatorletter\ldotp \Pred(\simulatorletter \link \idealfuncletter) \land T \equiv \behav{\simulatorletter \link \idealfuncletter} \}
\end{align*}
As with $\compHyper$, we again define a class of hyperproperties as the set of all of these hyperproperties:
$$\HypEqCls{\equiv}{\Pred} \isdef \{\hp \mid \exists \idealfuncletter\ldotp \hp = \HypEq{\equiv}{\Pred}(\idealfuncletter)\}.$$

This is precisely the class we are looking for.
When using the same equivalence and the source-language predicate~$\Preds$,
predicate-robustly preserving this class is equivalent to predicate-robustly preserving behavior up to~$\equiv$.

\begin{theorem}[\predrhc is \predrhp]
  \label{thm:predrhc-is-predrhpx}
  $$\def\arraystretch{1.35}
  \begin{array}{c}
    \vdash \comp{\cdot} : \predrhc(\equiv, \Predt, \Preds) \\
    \Updownarrow \\
    \vdash \comp{\cdot} : \predrhp(\HypEqCls{\equiv}{\Preds}, \Predt, \Preds)
  \end{array}$$
\end{theorem}
\noindent
This theorem is verified in Isabelle/HOL ($\mathtt{RHPXeqRRHC}$).

Note that, for all $\idealfuncletter$, $\compHyper(\idealfuncletter) = \HypEq{\compind}{\polyN}(\idealfuncletter)$,
meaning $\compHyper = \HypEqCls{\compind}{\polyN}$.
\Cref{thm:comprhc-ch} therefore follows as a special case of \Cref{thm:predrhc-is-predrhpx}
using \Cref{prop:comprhc-is-predrhc}.

\subsection{Connecting to \UC Security}

One obvious question this generalization raises is: how do these generalized notions of \RC correspond to \UC security?
Recall that the definitional structure of \UC security is extremely similar to the structure of \rhc (and \comprhc and \predrhc).
\UC security explicit demands environments that distinguish between executions while \predrhc allows arbitrary
equivalence relations over behaviors, but otherwise they are nearly identical.
Because the equivalence relation is arbitrary, we can consider the class of equivalences that consider
differences in behaviors between the source and target program for each environment separately, as we did in \Cref{sec:comp-rc}.

Taking this view, the predicates $\Preds$ and $\Predt$ represent restrictions on the behavior
of the protocols, ideal functionalities, attackers, and simulators,
and the equivalence $\equiv$ must specify any restrictions on environments as well as how similar the behaviors must be.
Setting the predicates to $\True$ and~$\equiv$ to~$=$, as we did above,
yields that requirement that behaviors must be identical, with no restriction on protocols, functionalities, attackers, simulators, or environments.
That definition corresponds precisely to perfect \UC security (\Cref{def:perf-uc}).
Since we have already shown the same assignments produce the definition of \rhc,
this correspondence immediately re-proves the result of \citet{uc-is-rc22}.

Similarly, if we use $\polysN$ and $\polytN$ as predicates and computational indistinguishability,
we recover a definition of computational \UC security (\Cref{def:comp-uc}),
and immediately recover \Cref{thm:compRHCeqUC}.

We are not, however, limited to these two cases.
For instance, consider using the trivial predicate $\True$ to leave protocols and attackers unrestricted,
and instantiating $\equiv$ with $\statind$, which leaves environments unrestricted but demands only \emph{indistinguishable} behaviors.
That is,
$$T_1 \statind T_2 \isdef \forall \env\ldotp T_1|_\env \indist T_2|_\env$$
The result is not perfect \UC security, as the behaviors may differ, but it is also not computational,
as there is no restriction on the complexity of any component of the system.
Instead, $\predrhc(\statind, \True, \True)$ corresponds to a third version of \UC security:
\emph{statistical} \UC security, where the distributions of behaviors must be statistically close, but not identical.

There are also other indistinguishability notions that appear in both the cryptography and language-based security literatures,
and this result can connect them to each other.
Examples from the cryptographic domain include
R\'{e}nyi divergence~\citep{baiImprovedSecurityProofs2018},
which is often used in lattice-based cryptography as it
provides a definition for when a search problem is computationally difficult,
Kullback-Leibler divergence~\citep{DBLP:conf/crypto/Micciancio017},
which has been used to to simplify proofs 
relating to fundamental definitions like adversarial advantage,
and more~\citep[e.g.][]{abboudCryptographicDivergencesNew2020}.

In language-based security, properties like noninterference~\citep{GoguenM82}
and observational determinism~\citep{McLean92,Roscoe95} are often defined
using an equivalence that erases certain (secret) parts of the trace and
demands the remaining (public) portion be identical~\citep{SabelfeldM03,ZdancewicM03,hyperproperties10}.
Different variations of these equivalences define different notions of noninterference
(e.g., termination-sensitive or insensitive, timing sensitive, or constant time),
or generalize to programs that include explicit declassifications or endorsements~\citep{ZdancewicM01,delimitedRelease03,nmifc17}.
Even other equivalence notions, like differential privacy~\citep{Dwork06}, may be possible,
particularly in combination with existing language-based results~\citep{lightdp17,shadowdp19,dpgen21}.

\begin{figure*}
  \centering
  \colorlet{derivediffcol}{RedViolet}
  \colorlet{specialcasecol}{YellowOrange}
  \tikzset{
    every node/.style={outer sep=2pt},
    derivediff/.style={myiff,dashed,color=derivediffcol},
    specialcase/.style={myiff,dotted,color=specialcasecol},
    pflabel/.style={font=\footnotesize,outer sep=0.33em},
  }
  \subfigure[Proof structure of \Cref{thm:predrhc-is-predrhpx} using arbitrary $\equiv$, $\Predt$, and $\Preds$.]{
    \begin{tikzpicture}
      \node (predrhp) {$\vdash \comp{\cdot} : \predrhp(\HypEqCls{\equiv}{\Preds}, \Predt, \Preds)$};
      \node[below=1.5em of predrhp] (emul-subset) {$\forall \prgs{}\ldotp \Emul{\equiv}{\Preds}{\Preds}(\prgs{}) \subseteq \Emul{\equiv}{\Predt}{\Preds}(\comp{\prgs{}})$};
      \node[left=3em of emul-subset] (emul-cont) {$\forall \prgs{}\ldotp \prgs{} \in \Emul{\equiv}{\Predt}{\Preds}(\comp{\prgs{}})$};
      \node[above=1.5em of emul-cont] (predrhc) {$\vdash \comp{\cdot} : \predrhc(\equiv, \Predt, \Preds)$};

      \draw[myiff] (emul-subset) -- node[above,pflabel] {\Cref{lem:emul-in-eq-contain}} (emul-cont);
      \draw[myiff] (emul-subset) -- node[right,pflabel] {\Cref{lem:emul-is-rhp}} (predrhp);
      \draw[myiff] (predrhc) -- node[left,pflabel] {\Cref{lem:emul-is-rhc}} (emul-cont);

      \draw[derivediff] (predrhc) -- node[above,pflabel] {\Cref{thm:predrhc-is-predrhpx}} (predrhp);
    \end{tikzpicture}
  }%
  \\[0.5\baselineskip]
  \subfigure[Proof structure of \Cref{thm:compRHCeqUC}.]{
    \begin{tikzpicture}
      \node (comp-emul) {$\forall \prgs{}\ldotp \prgs{} \in \Emul{\compind}{\polytN}{\polysN}(\comp{\prgs{}})$};
      \node[above=1.5em of comp-emul] (uc) {$\forall \prgs{}\ldotp \compuc{\comp{\prgs{}}}{\prgs{}}$};
      \node[right=2em of comp-emul] (predrhc-poly) {$\vdash \comp{\cdot} : \predrhc(\compind, \polytN, \polysN)$};
      \node[above=1.5em of predrhc-poly] (comprhc) {$\vdash \comp{\cdot} : \comprhc$};

      \draw[myiff] (comp-emul) -- node[left,pflabel] {\Cref{lem:emul-is-uc}} (uc);
      \draw[myiff] (predrhc-poly) -- node[right,pflabel] {\Cref{prop:comprhc-is-predrhc}} (comprhc);

      \draw[specialcase] (predrhc-poly) -- node[above,pflabel] {\Cref{lem:emul-is-rhc}} (comp-emul);

      \draw[derivediff] (uc) -- node[above,pflabel] {\Cref{thm:compRHCeqUC}} (comprhc);
    \end{tikzpicture}
  }
  \caption{Visual depiction of the proofs of our main theorems.
    We prove solid black implications directly, derive dotted the orange implication as a special case,
    and conclude dashed purple results.}
  \label{fig:proof-structure}
\end{figure*}
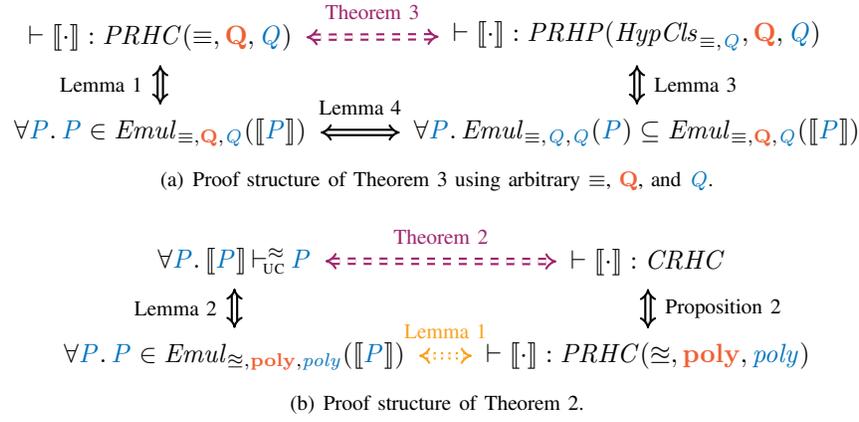

By inserting these equivalences into our framework, we immediately obtain a formal definition for the security they define
from both a robust compilation and a \UC standpoint.
The result is a meaningful way to relate more cryptographic security to secure compilation and more language-based security to cryptography.
Moreover, it provides the language needed to combine the two.
For example, some existing work combines noninterference-style definitions with computational hardness~\citep{FournetR08,FournetPR11},
and our work gives a way to situate those equivalences more broadly points to ways to extend and expand them.

Finally, the relation $\equiv$ does not need to be an equivalence relation.
Our proofs demand only that it is reflexive and transitive (i.e., a preorder), \emph{not} symmetric.
Indeed, instantiating $\equiv$ with $\subseteq$ produces a notion of behavioral refinement.

\section{Proof Approach}
\label{sec:proof-approach}

To prove \Cref{thm:comprhc-ch,thm:predrhc-is-predrhpx},
we develop an approach that connects directly to the definition of \UC security,
and allows a simple and direct proof of \Cref{thm:compRHCeqUC}.
Extending it with the generalization in \Cref{sec:general-equivs}
provides similar proof connecting \rhc to perfect \UC security,
simplifying and clarifying the proof by contradiction of \citet{uc-is-rc22}.

Our proofs rely on the idea of an \emph{emulation set}, which we denote $\EmulN(\prg{})$,
that represents the set of functionalities that $\prg{}$ securely emulates.
As in \Cref{sec:general-equivs}, we use a completely general definition of ``securely emulates''
and parameterize $\EmulN$ on an arbitrary equivalence~$\equiv$ and predicates~$\Pred$ and~$\Pred'$.
\begin{align*}
  & \Emul{\equiv}{\Pred}{\Pred'}(\prg{}) \isdef \\
  & \quad \{ \idealfuncletter \mid{} \forall \contextletter\ldotp \Pred(\contextletter \link \prg{}) \Longrightarrow
    \exists \simulatorletter\ldotp \Pred'(\simulatorletter \link \idealfuncletter) \\
  & \quad\phantom{\idealfuncletter\mid\forall\contextletter\ldotp{}}\quad \land \behav{\contextletter \link \prg{}} \equiv \behav{\simulatorletter \link \idealfuncletter} \}.
\end{align*}
This definition follows a very similar structure to \predrhc.
Setting~$\equiv$ to set equality~($=$) and both predicates to $\True$ again yields a definition of perfect emulation,
while setting~$\equiv$ to computational indistinguishability~($\compind$) and the predicates to $\polyN$ produces a definition of computational emulation.
The structure also allows for a very simple proof
that a compiler is \predrhc if and only if all compiled programs securely emulate their source program
(using the same predicates and equivalence).
\begin{lemma}[Emulation is \predrhc]\label{lem:emul-is-rhc}
  $$\vdash \comp{\cdot} : \predrhc(\equiv, \Predt, \Preds) \iff \forall \prgs{}\ldotp \prgs{} \in \Emul{\equiv}{\Predt}{\Preds}(\comp{\prgs{}})$$
\end{lemma}

Moreover, Axioms~\ref{ax:uc-rc} -- \ref{ax:final-bit} (\Cref{sec:comp-rc:uc}) directly connect this definition to \UC-style semantics.
For instance, the following results hold.
\begin{lemma}[Emulation is \UC]\label{lem:emul-is-uc}
  \begin{align*}
    \perfuc{\prot{}}{\idfu{}} & \iff \idfu{} \in \Emul{=}{\True}{\True}(\prot{}) \\
    \compuc{\prot{}}{\idfu{}} & \iff \idfu{} \in \Emul{\compind}{\polytN}{\polysN}(\prot{})
  \end{align*}
\end{lemma}

\Cref{lem:emul-is-rhc,lem:emul-is-uc} combine to provide direct proofs
for the perfect security result of \citet{uc-is-rc22} and \Cref{thm:compRHCeqUC}.

Emulation sets are also helpful in stating the class of hyperproperties a \predrhc compiler preserves,
making them useful in the proof of \Cref{thm:predrhc-is-predrhpx}.
In particular, a program~$\prg{}$'s behavior is in $\HypEq{\equiv}{\Pred'}(\idealfuncletter)$ in any context satisfying predicate~$\Pred$
precisely when $\idealfuncletter \in \Emul{\equiv}{\Pred}{\Pred'}(\prg{})$.
That is,
$$\def\arraystretch{1.2}
\begin{array}{c}
  \forall \ctxt{}\ldotp \Predt(\ctxt{}\linkt\prgt{}) \Longrightarrow \behavt{\ctxt{}\linkt\prgt{}} \in \HypEq{\equiv}{\Preds}(\idfu{}) \\
  \Updownarrow \\
  \idfu{} \in \Emul{\equiv}{\Predt}{\Preds}(\prgt{})
\end{array}$$
Applying this insight to \predrhp (\Cref{def:pred-rhpx}) yields the following result.
\begin{lemma}[Emululation specifies \predrhp]\label{lem:emul-is-rhp}
  $$\def\arraystretch{1.2}
  \begin{array}{c}
    \vdash \comp{\cdot} : \predrhp(\HypEqCls{\equiv}{\Preds}, \Predt, \Preds) \\
    \Updownarrow \\
    \forall \prgs{}\ldotp \Emul{\equiv}{\Preds}{\Preds}(\prgs{}) \subseteq \Emul{\equiv}{\Predt}{\Preds}(\comp{\prgs{}})
  \end{array}$$
\end{lemma}

The final hurdle to proving \Cref{thm:predrhc-is-predrhpx} is to notice that this subset relationship is equivalent to containment.
\begin{lemma}[Emulation Subset is Containment]\label{lem:emul-in-eq-contain}
  $$
    \src{\idealfuncletter} \in \Emul{\equiv}{\Predt}{\Preds}(\prgt{})
    \Longleftrightarrow
    \Emul{\equiv}{\Preds}{\Preds}(\src{\idealfuncletter}) \subseteq \Emul{\equiv}{\Predt}{\Preds}(\prgt{})
  $$
\end{lemma}

This result relies on the reflexivity of~$\equiv$ to show that $\src{\idealfuncletter} \in \Emul{\equiv}{\Preds}{\Preds}(\src{\idealfuncletter})$,
and the transitivity of~$\equiv$ to show that anything that $\src{\idealfuncletter}$~emulates $\prgt{}$~also emulates.
Notably, there is no need for~$\equiv$ to be symmetric.

The proofs of all theorems in \Cref{sec:comp-rc,sec:general-equivs},
which are verified in Isabelle/HOL, follow directly from the combination of these four lemmas using the structures shown in \Cref{fig:proof-structure}.

\section{A Mechanized proof of UC for Wireguard}\label{sec:wireguard}

We now describe how to leverage the results from \Cref{sec:comp-rc} into
a mechanized proof of \UC security for the \Wireguard protocol using the \Cryptoverif tool.
In particular, \Cref{thm:compRHCeqUC} means we can prove \UC security
by proving that a compiler that transforms the \Wireguard ideal functionality
into the \Wireguard protocol satisfies \comprhc.
We encode our notion of whole programs as \Cryptoverif games, and since \Cryptoverif is designed to analyze computational equivalence $\compind$ between two games, we can use it to verify \comprhc.

We begin with some background on \Cryptoverif and \Wireguard (\Cref{sec:cv,sec:wg})
before describing how we model \Wireguard in \Cryptoverif (\Cref{sec:wg-model-in-cv}).
Finally, we discuss the computational indistinguishability proof itself and limitations (\Cref{sec:ci-proof,sec:sub-disc}).

\subsection{The \Cryptoverif Tool}\label{sec:cv}

\Cryptoverif is a mechanized prover for properties of security protocols in the computational model. 
A \Cryptoverif proof is a polynomial-length sequence of cryptographic games---%
where an active attacker attempts to break some security property and is allowed to run multiple sessions in parallel.
Each game is represented in a language with a probabilistic semantics where all computations run in polynomial time,
and the goal is to show that each game is indistinguishable from the next.
\Cryptoverif proofs are thus proofs of computational indistinguishability between two games,
one of which is generally trivially secure (e.g., distinguishing between two encryptions of 0).

We use the game language of \Cryptoverif as an instance for both the source and target languages of the compilers described in \Cref{sec:comp-rc}.
Those languages require semantics that are probabilistic and where processes run in polynomial time, assumptions which \Cryptoverif satisfies.
Our strategy is to encode both the ideal functionality and the protocol as games, and prove that they are computationally indistinguishable.
This, in turn, shows that the compiler that translates the functionality into the protocol (and does nothing else) satisfies \comprhc,
and thus the protocol \UC-realises the functionality.

\Cryptoverif provides two input languages, channels and oracles.
We use the oracle language, as it is closer to \Cryptoverif's internal representation and thus grants more flexibility in the encoding.
In this model, the parties to a protocol are represented by oracles, which are simple probabilistic programs with shared state.
The adversary controls the network as well as the scheduling of messages by invoking these oracles.
An adversary invoking an oracle models sending a message to a party and then receiving a response.
This structure supports the attacker injecting messages---by invoking an oracle with an input it generated---%
dropping messages---by declining to send the output of one oracle as the input to another---%
or simply observing all network traffic.
\Cryptoverif models parties that answer many requests in a uniform way,
such as servers, by replicating an oracle:
given some finite bound~$n$, the attacker can use~$n$ different copies of the oracle,
each with their own state.

\subsection{The \Wireguard Protocol}\label{sec:wg}

\Wireguard is a widely-used protocol for establishing a VPN tunnel between two remote hosts in order to securely encapsulate all Internet Protocol (IP) traffic between them.
It consists of two parts: a key exchange and a record subprotocol.
To establish a tunnel, the key exchange subprotocol requires
the IP address and long-term public key for the remote host.
It then uses an instantiation of the Noise framework for key exchange
with fast, modern cryptographic primitives, like
Curve25519 and BLAKE2, to establishes two ephemeral keys (one per
direction) that parties use to
encrypt and authenticated messages in the subsequent record subprotocol.
The record subprotocol can then use these symmetric keys to construct a secure channel.

We verify a model of \Wireguard introduced by \citet{lippMechanisedCryptographicProof2019}, which is already encoded in \Cryptoverif.
\footnote{
  They provide several models. We build on a model in which long-term keys can be dynamically corrupted
  (\texttt{WG.25519.AB-BA.S\_i\_compr.S\_r\_compr.replay\_prot.cv}).
}
For the sake of simplicity, we focus on the 2-party version of \Wireguard, with a sender and receiver who are both honest throughout the run.

We also only model the record protocol, where payloads are transmitted after parties have agreed on a common, ephemeral symmetric session key.
The compositional nature of \UC security means that one could obtain a proof of the full \Wireguard protocol
simply by combining our proof with a corresponding proof for the key exchange protocol, which we leave for future work.
We make this choice due to a conceptual challenge in proving the security of key exchange.
The key confirmation message in \Wireguard---as well as other secure channel protocols like TLS---%
leaks a very small amount of information about the key, complicating composition arguments.
\citet{fischlinKeyConfirmationKey2016} discuss this problem and
provide a (non-composable) solution,
while the model of \citet[p.\ 242]{lippMechanisedCryptographicProof2019} omits this message altogether.

Finally, we introduce a modification to the original model. 
As an argument to the protocol, \citet{lippMechanisedCryptographicProof2019} include a secret bit that the two parties share.
Before sending a payload, a party waits for two messages from the adversary and uses the secret bit to determine which one to transmit.
The attacker must then determine the value of this secret bit.
This modeling choice deviates from any actual \Wireguard implementation, which should not include this secret bit or a second payload.
Instead, a more faithful model would take this direction from the environment, not the adversary.
To represent this more realistic model, we build a simple model of communication between the \Wireguard parties (sender and receiver),
add a dummy adversary, and use the \Cryptoverif attacker to model the environment.
We then add functionality from \citeauthor{lippMechanisedCryptographicProof2019}'s model in a piece-wise fashion until we have recovered the full record protocol.

\subsection{Modeling \Wireguard in \Cryptoverif}
\label{sec:wg-model-in-cv}

\paragraph*{Model Structure}
Recall that \Cryptoverif proves computational indistinguishability between two cryptographic games,
one representing the ``real world,'' which we denote $\ctxtdummy\linkt\prgt{WG}$,
and one representing the ``ideal world,'' which we denote $\simu{WG} \links \prgs{WG}$.
The real world consists of the sender and the receiver (both in $\prgt{WG}$) and the dummy attacker ($\ctxtdummy{}$).
The ideal world consists of the sender and receiver's functionalities ($\prgs{WG}$) and a simulator ($\simu{WG}$) that we devise to carry out the \UC proof.
Both the real and the ideal worlds interact with the same environment.
In our model, the role of the environment is taken by the \Cryptoverif attacker.

\paragraph*{The Simulator}
The simulator plays a key role in \UC proofs, since it determines the success of the proof itself.
Our simulator is inspired by simulators used in \UC proofs of encryption functionalities.
As the simulator performs multiple input/output steps, it is split over multiple oracles.
The oracle below is the part of the simulator that receives a cyphertext from the
environment and sends it to the functionality. 
\begin{center}
  \hspace*{6pt}\cprotect\fbox{%
\begin{lstlisting}
let Sim() =
 foreach i_Nrr <= n_M do ((*\label{lst:sim:li:foreach}*)
  Oe2aR (xc:bitstring, xn:counter_t) :=(*\label{lst:sim:li:oracle}*)
    get rcvd_counters(xn) in yield
    else insert rcvd_counters(xn);
         get simtable(zm,tc,tn)(*\label{lst:sim:li:zm}*)
         suchthat tc = xc && tn = xn
         in run SenderFSim(xn)).
\end{lstlisting}}
\end{center}
Line~\ref{lst:sim:li:foreach} creates $n_M$ instances of the simulator, indexed by $i_{Nrr}$.
Line~\ref{lst:sim:li:oracle} declares $n_M$ many instances of the \lstinline{Oe2aR} oracle (\textbf{O}racles modelling communication from the \textbf{e}nvironment via the dummy \textbf{a}ttacker to the \textbf{R}eceiver), which together constitute the simulator.
Each \lstinline{Oe2aR} instance receives a message containing a bitstring \lstinline{xc} and a counter \lstinline{xn} from the environment.
Then it checks if the counter has been received in the past.
If it has, the simulator aborts (\lstinline{yield}), otherwise it registers the counter.
Finally, each oracle instance checks if it has already received the bitstring with the same counter, and continues as \lstinline{SenderFsim}, which forwards the bitstring to the sender part of the functionality.

\paragraph*{Using \Cryptoverif Oracles}
We use the oracle interface of \Cryptoverif to describe both games and the attacker.
Concretely, each \Cryptoverif game provides the attacker with a set of oracles that can be queried by the attacker.
\Cryptoverif does not provide private communication (e.g., private channels), and this affects both the protocols and the various parties encoded in the real and ideal games.

\paragraph*{Modeling Private Communication}

Private communication between protocol/functionality and attacker/simulator is necessary to ensure that the trace does not
immediately reveal whether a run is in the real world or ideal world.
In a process calculus, private channels can serve this purpose~\citep{blanchetEfficientCryptographicProtocol2001,chevalDEEPSECDecidingEquivalence2018}.
In a stateful language, shared state is sufficient, but it requires proper encoding~\citep{BartheGHB11,PetcherM15}.
\Cryptoverif does not provide private communication natively, so we must encode it.

Oracles are allowed to share state and access each others variables directly.
While unusual from a semantics perspective, this method is often preferred by cryptographers for its simplicity, so we rely on the exposure of oracle states to model private communication between them.
If oracles are replicated (as in the case here), their variables must be accessed as arrays.
For instance, \lstinline{zm[simId]} would access the variable
\lstinline{zm} defined on line~\ref{lst:sim:li:zm} of the copy of
the above \lstinline{Sim} oracle with ID \mbox{\lstinline{simId}.} 
Concretely, this means \emph{any} oracle with the following code snippet
could access \lstinline{zm}.
\begin{center}
  \cprotect\mbox{%
\begin{lstlisting}[numbers=none]
find simId <= n_M suchthat defined(zm[simId])
    then return(zm[simId]).
\end{lstlisting}}
\end{center}
Here, the construct \lstinline{find ... suchthat}
non-deterministically sets such an index matching
the specified requirements: \mbox{\lstinline{simId}~$\leq$~\lstinline{n_M}}, 
and oracle \lstinline{simId} has initialized \lstinline{zm}.

In the real world, the only necessary private communication is between the protocol ($\prgt{WG}$) and the dummy attacker ($\ctxtdummy{}$).
Since the dummy attacker $\ctxtdummy$ is just a proxy that forwards all messages between $\prgt{WG}$ and the environment, represented by the \Cryptoverif attacker,
we elide the dummy attacker and have $\prgt{WG}$ communicate directly with the environment / \Cryptoverif attacker.
For example, the \lstinline{Sender} directly returns to the \Cryptoverif attacker in \lstinline{Oe2S}, without calling the dummy attacker.
\begin{center}
  \hspace*{6pt}\cprotect\fbox{%
\begin{lstlisting}
let Sender(ks:key_t) =
  foreach i_Nis <= n_M do (
    Oe2S(m:bitstring, counter:counter_t) :=
      let preparemsgsucc(cipher_data:bitstring)
          = prepare_msg(m, counter, ks)
      in return(cipher_data)
  ).
\end{lstlisting}}
\end{center}

The ideal world requires more interesting private communication between the ideal functionality ($\prgs{WG}$) and the simulator ($\simu{WG}$).
However, $\simu{WG}$ is not stateless, so we need to manually inline the receiving party where it is called.
We carefully track names
to ensure that the receiving entity only accesses information that would be passed by communication.
Unfortunately, this manual encoding can be error prone, since \Cryptoverif allows the receiver to access any part of the sender state without a warning.

To illustrate this private communication, see the following \lstinline{SenderFSim} oracle, which accesses the variables~\lstinline{m}, \lstinline{counter}, and \lstinline{c} that are defined outside of its scope, in the oracle \lstinline{Sender} above.
\begin{center}
  \hspace*{6pt}\cprotect\fbox{%
\begin{lstlisting}
let SenderFSim(xn:counter_t) =
  find senderid <= n_M
    suchthat defined(m[senderid],
                     counter[senderid],
                     c[senderid])
      && counter[senderid] = xn
  then return(m[senderid]).
\end{lstlisting}}
\end{center}
\lstinline{SenderFSim} constitutes the adversary/simulator
interface of the sender part of the functionality (whereas \lstinline{Sender} constitutes the environment interface and sets \lstinline{m}).
Because we model multiple sessions of the overall protocol running in parallel,
the simulator may receive an out-of-order message from the environment,
so the simulator needs to identify the correct session of the
functionality. It does so by providing the (public) counter \lstinline{xn},
which \lstinline{SenderFSim} uses to search the correct (internal) session identifier.
First, the functionality finds the session, identified via \texttt{senderid}, that emitted the cyphertext.
The functionality searches for a session that (a) is terminated, i.e., one where
\lstinline{m[senderid]},
\lstinline{Fcounter[senderid]} and
\lstinline{c[senderid])} are defined,
and (b) where the oracle's counter 
\lstinline{Fcounter[senderid]} matches the counter 
\lstinline{xn} received from the simulator.
Up to now, we have simply encoded that 
the functionality, after producing its output to the simulator,
waits for another message that it
matches against 
\lstinline{xn}; this is necessary because oracles can only produce one
output per input.
Finally, the functionality returns
the message of the session now identified:
\lstinline{m[senderid]}.

The final model is around 200 lines of \Cryptoverif code.

\subsection{Computational Indistinguishability Proof}\label{sec:ci-proof}

We prove our two models (real and ideal) are computationally indistinguishable through a series
of eight \Cryptoverif games that the tool does not find automatically.
Two of those games are cryptographic reductions, while the others are perfect equivalences.
We guide \Cryptoverif to these games with a proof script consisting of seven steps.

The scrip starts from the real world $\ctxtdummy\linkt\prgt{WG}$.
The first step applies the IND-CTXT assumption of the encryption scheme, which guarantees ciphertext integrity.
This check matches a structural check in the ideal functionality and ensures that every received ciphertext was sent by the sender.
The second step applies the IND-CPA assumption of the encryption scheme, which guarantees ciphertext confidentiality.
This step replaces message encryptions with encryptions of zero.
The other five steps consist of removing dead code, inlining variables,
and minimally modifying code structure to produce exactly the ideal world $\simu{WG}\links\prgs{WG}$.

\subsection{Discussion}\label{sec:sub-disc}

The lack of private communication such as private channels led us to use cross-oracle state sharing to transfer information between oracles and to inline parties where they are called. 
This is a manual process that is highly prone to encoding errors, and we believe adding private channels to \Cryptoverif would mitigate this issue. 
Indeed, this feature seems available internally, but we confirmed through the manual and communication with the maintainers that is is not exposed to the user.
In the long term, the addition of a module system that allows for compositional proofs might be even better, as it could syntactically ensure that composition is correctly encoded, for example according to \citet[Axiom 5 to 7]{uc-is-rc22}.
In addition, such a system could easily validate that states are properly encapsulated, e.g., that simulator oracles cannot directly access the state of the functionality.

\section{Related Work}
\label{sec:rw}

The most closely-related work is that of \citet{uc-is-rc22}, which highlights the \UC--\RC connection but does not broach the topic of computational indistinguishability.
The authors use the connection to mechanize the proof that 1-bit commitment protocols~\citep{Canetti-UC01,ilc19} are \UC secure in the static as well as in the dynamic corruption cases.

\subsection{Universally Composable Cryptographic Models}

Universal composability~\citep{Canetti-UC01} extended
simulated-based notions of correctness and privacy for multi-party
computation~\citep{Goldreich:2006:FCV:1202577}
to the interactive case, additionally ensuring that security
guarantees are preserved when a protocol is used as a subroutine within
a larger protocol.
Many competing frameworks~\citep{backesReactiveSimulatabilityRSIM2007,
HofheinzS15, iuc19, iitm-uc20,
simpleUC16} followed to improve on perceived
inadequacies in the model or to correct subtle errors in the proofs.
Currently, the original \UC framework is in its sixth version. A common
point of contention is how to define polynomial runtime in a reactive
system. 
\citet{HofheinzUM13} explain the
problem in great detail and propose
a solution that is (conceptually) reflected in all these frameworks
\citep{Canetti-UC01,HofheinzS15,
iuc19, iitm-uc20, simpleUC16},
at least in their latest versions.
Our abstract poly-time predicate over attacker-protocol combinations
(see \Cref{sec:comp-rc}) can capture the poly-time notions in each of
them.

These frameworks define \UC bottom-up, fixing a plethora of
technical details like the machine model (often interactive Turing
Machines), message formats, message and process schedulers,
virtualisation mechanisms, addressing mechanisms, etc.
None of these details are at all similar to how networks and programs operate in the real world,
yet many of the differences between the frameworks boil down to technical
minutia at very low-level of abstraction.
This is reflected in the proofs, which are as hard to formalize as
they are to write down. 

By contrast, our results operate at a much higher level of abstraction.
\citet{uc-is-rc22} provide a high-level composition proof for this model,
which we verified in Isabelle/HOL is not impacted by any of our modifications ($\mathtt{composition}$).
Indeed, that result relies on a small set of assumptions \citep[Axioms 5 to 7]{uc-is-rc22}
that most likely hold for any of the frameworks discussed above.

\citet*{easyuc19} confirmed these issues, when
they
mechanized parts of UC in
EasyCrypt~\citep{BartheGHB11}, lamenting
that
``despite the relative simplicity [of their case studies], [proving UC] took an immense amount of work.''
It is thus unsurprising that other attempts to mechanize the
concept~\citep{AbateHR+21, BasinLMS21, ilc19}
avoid this technical baggage and
build their frameworks on formal languages that provide a much higher degree of
abstraction. Still, all these frameworks perceive \UC as a property
between programs written in one and the same language.
With \Cref{thm:compRHCeqUC}, one can use \RC as a tool
to transfer such properties across language boundaries.

\subsection{Alternative Tools for \Cryptoverif}

We verified our case study in \Cryptoverif, but there are a variety of other tools for cryptographic verification.

EasyCrypt~\citep{easycrypt} is a stand-alone theorem prover with focus on cryptographic
primitives and small protocols.
It reasons at at a code level using probabilistic relational Hoare logic and external SMT solvers.
There are also various embeddings of probabilistic languages in general-purpose theorem provers like
CryptHOL~\citep{BasinLMS21} and FCF~\citep{PetcherM15}.
They all require most of the proof to be written out, although some
automated proof tactics can help with simple steps.

Squirrel~\citep{baelde:hal-03172119} is a protocol-specific prover
for computational indistinguishability based on a computationally-sound
attacker model that can be symbolically reasoned about. In contrast to
\Cryptoverif and EasyCrypt, which focus on program transformations to deduce
equivalences, Squirrel's reasoning focusses on (symbolic) traces.
Currently Squirrel proofs are as detailed as proofs in EasyCrypt and similar tools,
though the project aims to reduce the complexity.

Only EasyCrypt has previously been used for \UC-style proofs~\citep{easyuc19}.
Leveraging our results, we were able to produce a proof in \Cryptoverif,
which we chose because it promised a higher degree of automation.
Our case study confirms this promise, though manual guidance was still required.

\subsection{Language-Based Tools for Cryptographic Security}

Several language-based tools exist with the goal of easing the process of developing and verifying secure cryptographic protocols.
Languages like Wysteria~\citep{wysteria14}, Wys$^*$~\citep{wysstar18} and Symphony~\citep{symphony23} are designed to simplify
the design and implementation of secure distributed systems using cryptography.
Viaduct~\citep{viaduct21} and Jif/Split~\citep{jifSplit02,ZhengCMZ03}
use security policies specified by information flow labels to automatically partition programs,
synthesizing uses of particular cryptographic primitives when necessary.
These systems generally do not contain formal proofs of security,
and we hope our results will make it easier to provide them with \UC-style guarantees.

Security type systems can also help prove security.
An information-flow type system can ensure proper combination of specific cryptographic primitives~\citep{FournetR08,FournetPR11}.
Owl~\citep{GancherGSDP23} uses a security type system with built-in primitives
to enforce computational guarantees.
While these builtins do not fit our goal of
abstracting (any) primitive via functionalities,
they provide a meaningful level of composition
and are worth investigating. In general, type-systems are fast
and compositional.

\subsection{Robust Compilation}
\citet{journey-beyond19} introduced a large hierarchy of \RC criteria
that have been used to reason about the security of compilers that preserve memory safety~\citep{PatrignaniGarg21,AbateAB+18},
absence of speculation leaks~\citep{PatrignaniG21}, and cryptographic constant time~\citep{KolosickSC+23,KruseBP23}.
They require source and target languages to have the same trace model, though they show how to lift that limitation in subsequent work~\citep{AbateBC+21}.
While we also require our languages to share a trace model, we make that choice for simplicity.
The same approach should generalize our result as well.
Interestingly, lifting the same limitation in \UC would let us formalize the existence of two distinct, but related, environments, one in the real and one in the ideal world.
We are unaware of a notion of \UC with distinct environments; after all, they are all just (interactive) Turing Machines.

\section{Conclusion}
\label{sec:conc}

In this work, we have generalized the connection between \UC and \RC to the computational setting,
and then generalized the connection further to arbitrary indistinguishability relations and predicates over programs.
We showcased the benefits of this expanded connection by using \Cryptoverif, a tool for proving computational indistinguishability,
to mechanize a proof of \UC security for the \Wireguard protocol.

These results let us conclude that $\Cryptoverif$ can be used to
modularize cryptographic proofs, which are currently very monolithic.
We believe our demonstrated connections will extend similar benefits to other tools,
though we leave this investigation for future work.

Additionally, there exists a compiler from \Cryptoverif into OCaml~\citep{cadeProvedGenerationImplementations2015}
that the authors proved, already in 2013, robustly preserves hyperproperties.
Though the \RC framework was only developed years later, the result is equivalent.
Since translating from the ideal functionality into the protocol is a compiler,
we can create a toolchain that translates \Cryptoverif ideal functionalities all the way to executable OCaml protocols.
This toolchain would combine properties of sequential composition of compilers~\citep{KruseBP23} and our results
to ensure end-to-end guarantees of \UC security for real-world executable protocols.
However, as we hint at the end of \Cref{sec:wireguard}, a module system in \Cryptoverif would streamline the creation of this toolchain, which we also leave for future work.

\section*{Acknowledgments}

Thanks to the anonymous reviewers for their insightful comments and suggestions.
This work was partially supported by a gift from:
the Italian Ministry of Education through funding for the Rita Levi Montalcini grant (call of 2019).

\bibliography{../ethan,../ephemeral,../RCisUC}

\end{document}